\documentclass[pre,showpacs,twocolumn,amsmath,amssymb,superscriptaddress,floatfix]{revtex4}

\usepackage{graphicx}

\usepackage{bm}

\usepackage{latexsym}

\begin{document}


\title{Fluctuations and Correlations in Lattice Models for Predator-Prey
       Interaction}

\author{Mauro Mobilia} \email{mauro.mobilia@physik.lmu.de}
\affiliation{Arnold Sommerfeld Center for Theoretical Physics and CeNS,
     	Department of Physics, Ludwig-Maximilians-Universit\"at M\"unchen,
	D-80333 Munich, Germany}
\affiliation{Department of Physics and
     	Center for Stochastic Processes in Science and Engineering,
     	Virginia Polytechnic Institute and State University,
     	Blacksburg, Virginia 24061-0435, U.S.A.}
\author{Ivan T. Georgiev} 
\affiliation{Integrated Finance Limited, 630 Fifth Avenue, Suite 450,
        New York, NY 10111, U.S.A.}
\affiliation{Department of Physics and
        Center for Stochastic Processes in Science and Engineering,
        Virginia Polytechnic Institute and State University,
        Blacksburg, Virginia 24061-0435, U.S.A.}
\author{Uwe C. T\"auber}  
\affiliation{Department of Physics and
        Center for Stochastic Processes in Science and Engineering,
        Virginia Polytechnic Institute and State University,
        Blacksburg, Virginia 24061-0435, U.S.A.}

\begin{abstract}
Including spatial structure and stochastic noise invalidates the classical
Lotka--Volterra picture of stable regular population cycles emerging in models
for predator-prey interactions.
Growth-limiting terms for the prey induce a continuous extinction threshold
for the predator population whose critical properties are in the directed 
percolation universality class.
Here, we discuss the robustness of this scenario by considering an ecologically
inspired stochastic lattice predator-prey model variant where the predation 
process includes next-nearest-neighbor interactions.
We find that the corresponding stochastic model reproduces the above scenario
in dimensions $1< d \leq  4$, in contrast with mean-field theory which predicts
a first-order phase transition.
However, the mean-field features are recovered upon allowing for 
nearest-neighbor particle exchange processes, provided these are sufficiently
fast.
\end{abstract}

\pacs{87.23.Cc, 02.50.Ey, 05.40.-a, 05.70.Fh}

\date{\today}

\maketitle

In 1920 and 1926, respectively, Lotka \cite{Lotka} and Volterra \cite{Volterra}
devised a simple coupled set of differential equations to describe an
autocatalytic reaction model and the statistics of fish catches in the 
Adriatic.
The Lotka--Volterra model (LVM) has since become one of the central paradigms
for the emergence of periodic oscillations in nonlinear systems with competing
constituents \cite{Haken}, and features prominently in textbooks, from
undergraduate-level population biology \cite{Neal} to ecology
\cite{May,Maynard} and mathematical biology \cite{Murray} as, for instance,  
it can also be formulated as a host-pathogen model \cite{host}.
Yet it has often been severely criticized as being biologically unrealistic and
mathematically unstable \cite{Neal,Murray,Dunbar}.
Recent investigations of zero-dimensional \cite{McKane} and spatial stochastic
models \cite{Matsuda,Durrett,Provata,Lipowska,Albano,host} have shown that
this criticism definitely pertains to the original deterministic rate
equations; however, it turns out that the stochastic, or lattice, two-species predator--prey
model variants display quite robust properties, rather insensitive on the
details of the underlying microscopic processes (for a recent overview, see 
Ref.~\cite{Mobilia}).
In particular, the lattice predator-prey models (LPPM), display the following features:
(i) The population densities typically display erratic (rather than regular
periodic) oscillations, with amplitudes that vanish in the thermodynamic limit
\cite{Provata}, caused by persistent and recurrent
predator--prey activity waves that form complex spatio-temporal structures
\cite{Boccara};
(ii) when the prey population growth is limited (finite carrying capacity,
local site restrictions), there exists an extinction threshold for the predator
population \cite{Lipowska,Albano}; this constitutes a nonequilibrium 
active-to-absorbing-state phase transition with the critical exponents of 
{\em directed percolation} (DP) \cite{Hinrichsen,Grassberger}. 
Also, for host-pathogen models with two types of pathogens, the invasion of the
system by one pathogen (the other becoming extinct) through oscillatory 
behavior, was reported using mean-field and pair-approximation treatments 
\cite{host}.

As noted by various authors \cite{Provata,Lipowska,Albano,Boccara}, a more
realistic description of the predator-prey interaction should include the
possibility for the agents to move.
In fact, in real ecosystems prey tend to evade the predators, while the
predators aim to pursue the prey.
One approach to account for the motion of the agents is to consider
{\em diffusion} (nearest-neighbor hopping) of predators and/or prey, which
however does not really affect the global properties of the LPPM
\cite{Lipowska}.
Another approach, to be considered here, is to assume a nearest-neighbor
exchange process (among any two agents: predators, prey and empty sites) in
the following referred to as `stirring'.
It has to be noticed that both diffusion and stirring processes are not taken
into account at the (mean-field) rate equation level.
In addition, some recent investigations have included long-range processes in
two-dimensional LPPM, reporting quite different results on the existence
\cite{Albano} or absence \cite{Lipowska} of `self-sustained oscillations' (in
the thermodynamic limit). 
We also notice that  for spatial host-pathogen models (with nearest-neighbor 
interactions), the rate equations include algebraic nonlinear terms of power 
$2d+1$ in dimensions $d$ \cite{host}. 
Thus, an understanding of the joint effect of long-range interactions and of
the agents' motion is desirable and relevant from ecological and statistical 
physics points of view.

In this Rapid Communication, we aim to shed further light on the remarkable
robustness of the LPPM scenario.
To this end, we study an ecologically inspired stochastic lattice
predator--prey model with {\em next-nearest-neighbor} interaction
(NNN-LPPM), both in the presence and absence of a {\em nearest-neighbor
exchange process} (`stirring').
We will demonstrate a subtle interplay between the correlations generated by
the NNN interaction and the stirring process.
As a result, there is a regime where the NNN-LPPM phase diagram indeed follows
the LPPM scenario outlined above, with a {\em continuous} predator extinction
transition in the DP universality class.
On the other hand, we shall also see under which unexpected conditions a
{\em first-order} phase transition can occur as a consequence of the
competition between the short-range exchange and the NNN predator--prey
interactions.

To begin, we outline the main properties of the deterministic LVM and then of
the corresponding LPPM \cite{Haken,Neal,Murray}.
Consider two chemical species subject to the reactions $A \to \oslash$ (decay
rate $\mu>0$), $B +\oslash \to B + B$ (branching rate $\sigma>0$), and $A + B \to A + A$
(predation rate $\lambda>0$).
Neglecting any spatial variations and fluctuations of the concentrations
$a({\bm x},t)$ and $b({\bm x},t)$ of `predators' $A$ and `prey' $B$, one
obtains the classical LVM rate equations:
$\dot{a}(t) = \lambda \, a(t) \, b(t) - \mu \, a(t)$ and
$\dot{b}(t) = \sigma \, b(t) - \lambda \, a(t) \, b(t)$.
These  deterministic equations have as stationary states $(a^*,b^*) = (0,0)$ 
(extinction), $(0,\infty)$ (predators extinct, prey proliferation), and 
$(a^{*}_c,b^{*}_c) = (\sigma/\lambda , \mu/\lambda)$ (species coexistence). 
The unstable fixed points $(0,0)$ and $(0,\infty)$ constitute {\em absorbing} 
states of the dynamics.
The existence of a conserved first integral of the deterministic rate 
equations, $K(t) = \lambda [a(t) + b(t)] - \sigma \ln{a(t)} - \mu \ln{b(t)} 
= \text{constant}$, implies oscillatory kinetics around $(a^{*}_c,b^{*}_c)$.

Since this center singularity is {\em unstable} with respect to introducing
model modifications \cite{May,Murray}, the LVM rate equations are often
rendered more `realistic' by introducing growth-limiting terms
\cite{Neal,Murray}.
For the LVM, this amounts to replacing the rate equation for species $B$ with
$\dot{b}(t) = \sigma \, b(t) \left[ 1 - \rho^{-1} \, b(t) \right]
- \lambda \, a(t) \, b(t)$ ($\rho$ is the prey
`carrying capacity'; growth-limiting terms for the predators do not
induce any qualitative changes.)
The three fixed points are now shifted to
$(a^*,b^*) = (0,0)$ (extinction), $(0,\rho)$ (predators extinct, system
saturated with prey), and $(a^{*}_r,b^{*}_r)$ with
$a^{*}_r = \left( 1 - \mu / \lambda \, \rho \right) \sigma /\lambda $,
which is in the physical region ($0 < a^{*}_r \leq 1$) if $\lambda > \mu / \rho$,
and $b^{*}_r = \mu / \lambda$.
Linear stability analysis reveals $(0,0)$ to be a saddle-point, whereas
$(0,\rho)$ is stable (node) if $\lambda < \mu / \rho$ (when $a^{*}_r < 0$), and a
saddle-point (stable in the $b$ direction) otherwise.
When $\lambda > \mu / \rho$, the coexistence state $(a^{*}_r,b^{*}_r)$ is stable; it is
either a node or a focus, associated with spiral trajectories in the $(a,b)$
phase plane \cite{Murray}. 
Thus, at the rate equations level, $\lambda_c = \mu / \rho$ is the critical 
predation rate.
The global stability of $(a^{*}_r,b^{*}_r)$ is established by the existence of a
Lyapunov function ${\cal L}(a,b)= \lambda\{a^{*}_r\ln{a(t)}+b^{*}_r\ln{b(t)} -a(t)-b(t)\}$ \cite{Murray}.
Many of these features re-emerge in  {\em stochastic} LPPM with site restriction.
Indeed, Monte Carlo simulations \cite{Matsuda,Provata} yield that as in 
mean-field theory the coexistence fixed point is either a node or a focus.
In the latter case,  amazingly rich spatio-temporal patterns of persistent 
predator--prey `pursuit and evasion' waves \cite{Murray,Dunbar} emerge, 
inducing erratic correlated population density oscillations. 
In {\em finite} systems, these quasi-periodic fluctuations appear on a global 
scale, but the amplitude of the density oscillations decreases with system size
\cite{Provata}.
A completely different picture emerges when the active fixed point is a node,
just above the predators' extinction threshold $\lambda_c$:
Instead of the intricate front patterns, small predator `clouds' effectively
diffuse in a sea of prey \cite{Mobilia}.
If the value of $\lambda$ is reduced further (keeping the other rates fixed), 
at the critical value $\lambda_c$ the system reaches the absorbing state. 
This active-to-absorbing phase transition is found to be in the directed 
percolation (DP) universality class \cite{Grassberger}; this is also true for 
many LPPM variants \cite{Lipowska,Provata,Albano}. 
These results can be understood from the master equation: For the above
reactions one may derive an equivalent field theory action \cite{Howard}, which
near $\lambda_c$ can be mapped onto Reggeon field theory \cite{Mobilia}, known
to describe the asymptotic DP scaling laws \cite{Grassberger, Howard,Janssen}.

In most LPPM (see, e.g., Refs.~\cite{Lipowska,Provata}), the `predation' 
process subsumes \emph{nearest-neighbor} interaction and the effects on both the prey 
and the predators in a \emph{single} `reaction'.
More realistically, one should split this into two processes, and thereby 
introduce {\em two} independent time scales.
This leads to the following stochastic reaction scheme that incorporates a {\em three-site} (NNN) process:
(a) A predator reproduces in the vicinity of a prey ({\em favorable} environment)
according to the triplet reaction $A + \oslash + B \to A + A + B$ [with rate $\delta/z(z-1)$; $z=2d$ is the 
coordination number of a $d$-dimensional hypercube];
(b) a predator consumes a neighboring prey (rate $\eta / z$), leaving an
empty site, according to the binary process $A + B \to \oslash + A$;
(c) we shall also allow for an efficient {\em mixing} process, through
particle exchange with rate ${\cal D} / z$ (`stirring') between two
{\em neighboring} sites regardless of their content \cite{Durrett}.
Besides these reactions, we still consider the processes
$B+\oslash\to B+B$ (rate $\sigma/z$) and $A\to \oslash$ (rate $\mu$).
Assuming full site restriction, i.e. allowing at most one particle per site,
the mean-field (MF) rate equations now read
\begin{eqnarray}
\label{NNN_MFa}
&&\dot{a}(t) = \delta \, a(t) \, b(t) \, [1 - a(t) - b(t)] - \mu \, a(t) \\
\label{NNN_MFb}
&&\dot{b}(t) = \sigma \, b(t) \, [1 - a(t) - b(t)] - \eta \, a(t) \, b(t) \ .
\end{eqnarray}
These equations can be obtained from the master equation of 
our NNN-LPPM upon factorizing the three-point correlators as products 
of the corresponding densities $a, b$.
In contrast with the LVM, the nonlinear term in Eq.~(\ref{NNN_MFa}) is
{\em cubic} (NNN interaction); the site restriction appears through the
growth-limiting factors $1 - a - b$.
Note that the mixing parameter ${\cal D}$ does not enter the rate equations
(but would appear in the equations for the three-point and higher
correlation functions).
Equations~(\ref{NNN_MFa}), (\ref{NNN_MFb}) admit {\em four} fixed points,
provided $\delta > \delta_c = 4 \mu (\sigma + \eta) / \eta$.
In addition to the previous absorbing states, $(a^*,b^*) = (0,0)$ and $(0,1)$,
the two new nontrivial steady states $(k=1,2)$ are given by
\begin{eqnarray}
\label{FPNNN1}
a^*_k = \frac{\sigma}{2 (\sigma + \eta)}
\left[ 1 - (-1)^k \sqrt{1 - \frac{\delta_c}{\delta}} \right]
\end{eqnarray}
and $b^*_k =\frac{1}{2}[1+(-1)^{k}\sqrt{1-\delta_c/\delta}]$.
\begin{figure}[!t]
\includegraphics[angle=0,width=1.58in]{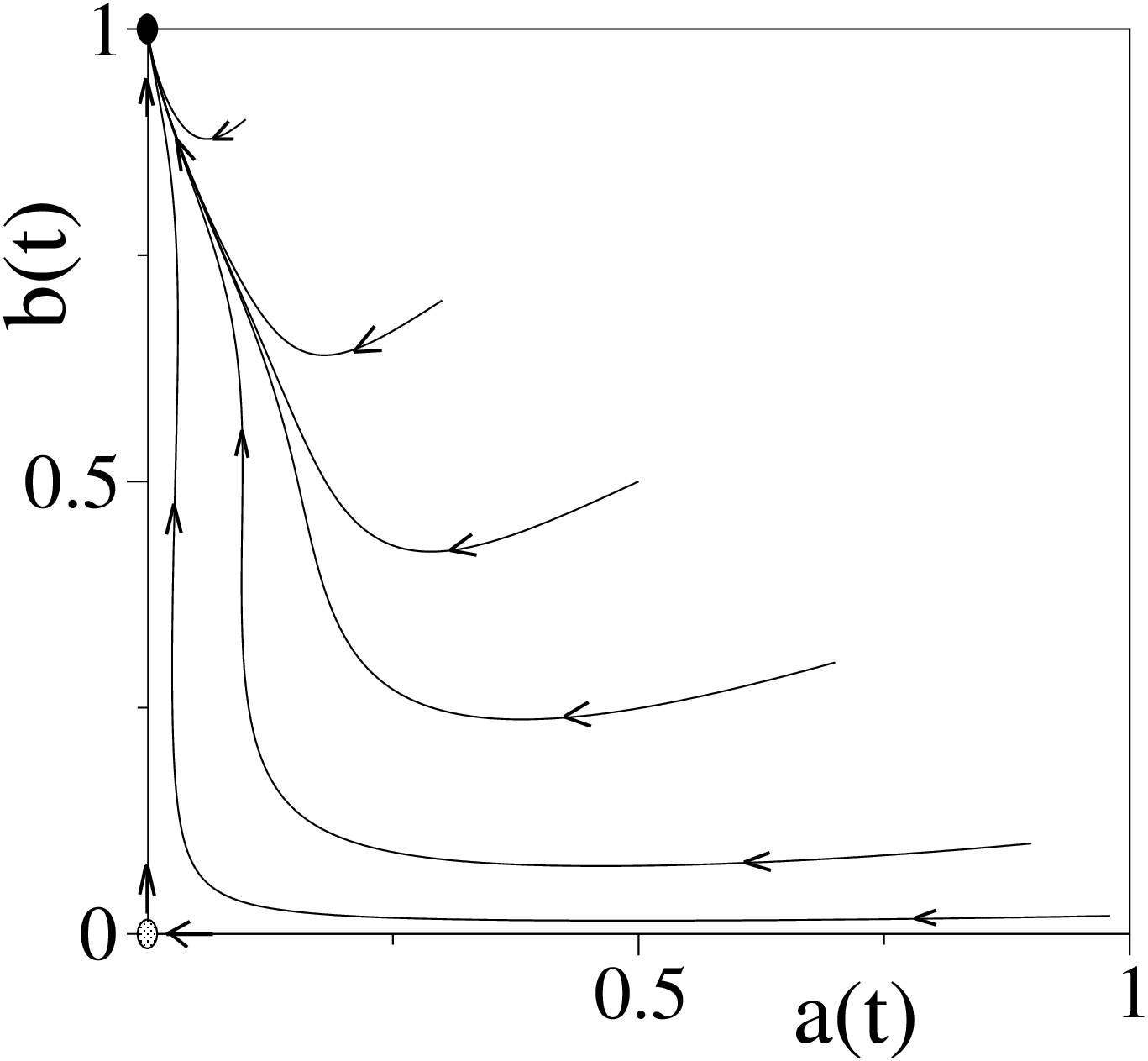}
\includegraphics[angle=0,width=1.51in]{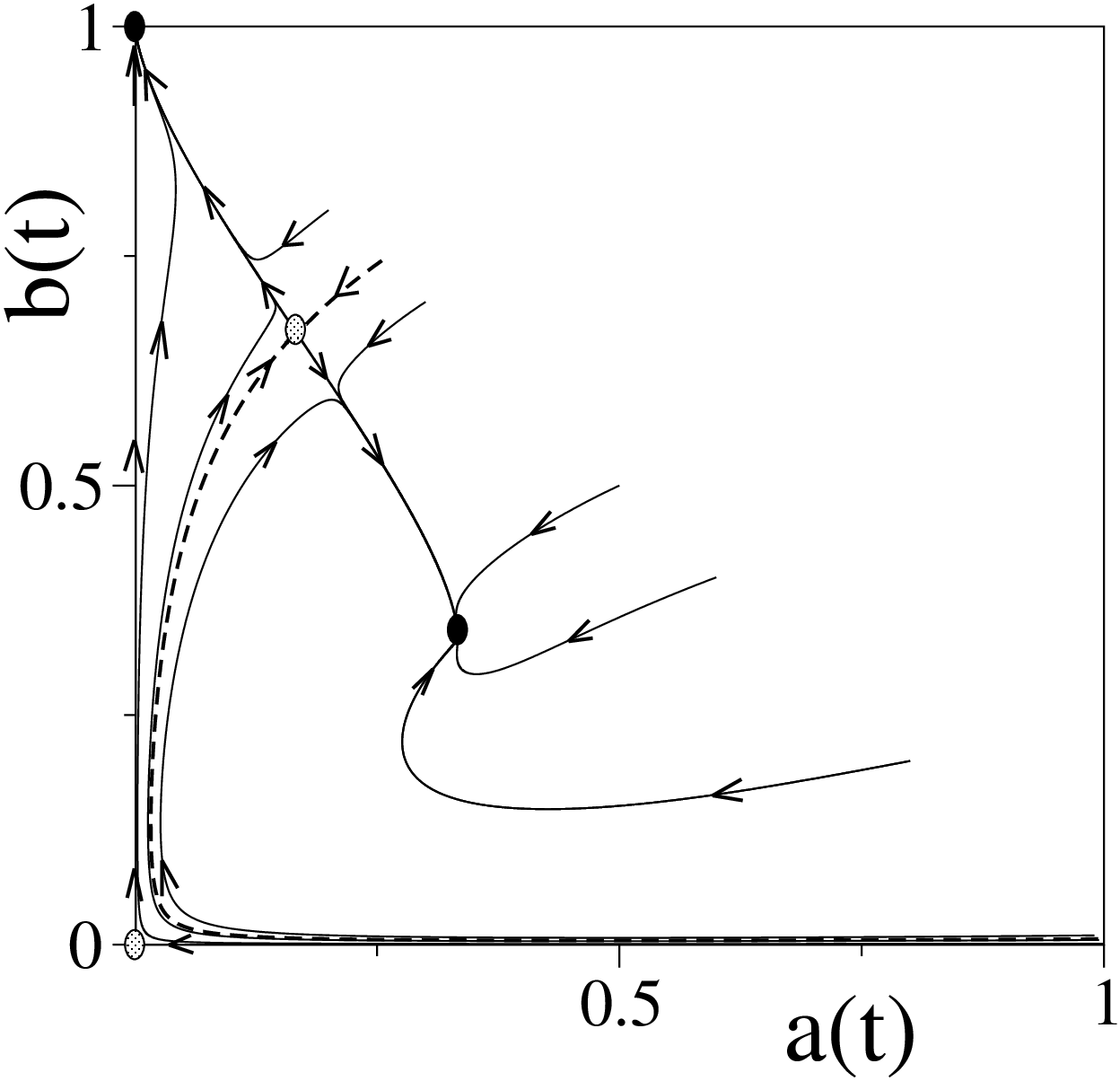}
\caption{Flows in the phase plane from integrating
Eqs.~(\ref{NNN_MFa}), (\ref{NNN_MFb}) for $\mu = \sigma = \eta = 1$, with
$\delta = 4$ (left) and $\delta = 9$ (right). Left: $(0,1)$ is the only stable 
fixed point (node). Right: there is an additional stable (node) active fixed 
point $(a_1^*,b_1^*) = (1/3,1/3)$; while $(0,0)$ and 
$(a_2^*,b_2^*) = (1/6,2/3)$ are unstable. The slopes of the separatrices at 
$(a_2^*,b_2^*)$ are $\approx 1.126 $ (dashed line) and $\approx -1.568 $ 
-(see text).}
\label{portrait}
\end{figure}
These active fixed points $(a^*_{1,2} , b^*_{1,2})$ correspond to two
{\em distinct} predator--prey coexistence phases. From linear stability 
analysis we infer that the absorbing state $(0,1)$ is {\em always} a stable 
node: the associated Jacobian eigenvalues read $\epsilon_+(0,1) = -\mu$ [with 
eigenvector ${\bm v}_+ = \left( \{\mu-\sigma\}/\{\sigma + \eta\},1 \right)$] 
and $\epsilon_-(0,1) = -\sigma$ [eigenvector ${\bm v}_- = \left( 0,1\right)$].
On the other hand, $(0,0)$ is an unstable saddle-point, with eigenvalues
$\epsilon_+(0,0) = \sigma$ [with unstable eigendirection ${\bm v}_+ = (0,1)$]
and $\epsilon_-(0,1) = -\mu$ [stable eigendirection ${\bm v}_- = (1,0)$].
Without loss of generality, we just discuss the stability of the active fixed 
points (\ref{FPNNN1}) when $\eta = \mu = \sigma = 1$.
In this case, $\delta_c = 8$ and the eigenvalues of the Jacobian respectively
read ($k=1,2$): $\epsilon_\pm(a_k^*,b_k^*) = -\frac{1}{4} \left[3+ (-1)^k
\sqrt{1 - 8 / \delta}\right] \pm \frac{1}{4} \sqrt{22 (-1)^k \,
\sqrt{1 - 8 / \delta} + 10 - 8 / \delta}$.
Thus, the active fixed point $(a_1^*,b_1^*)$ is stable,
$\Re(\epsilon_\pm(a_1^*,b_1^*)) < 0$, while $(a_2^*,b_2^*)$ is a saddle-point.
More generally, for fixed $\mu, \sigma, \eta$ there exists a value
$\delta_s > \delta_c$ such that $(a_1^*,b_1^*)$ is a stable node if
$\delta_c < \delta \leq \delta_s$, and a stable focus, i.e.,
$\Im(\epsilon_\pm(a_1^*,b_1^*)) \neq 0$, if $\delta > \delta_s$.
When $\eta = \mu = \sigma = 1$, $\delta_s = (29+11\sqrt{7})/6 \approx 9.68388$.
Typical phase portraits as predicted by Eqs.~(\ref{NNN_MFa}), (\ref{NNN_MFb})
are illustrated in Fig.~\ref{portrait}.
The eigenvectors associated with $(a_2^*,b_2^*)$ give the slope of the
separatrices in its vicinity: $4(\sqrt{\delta}-\sqrt{\delta-8}) / \Bigl(
\sqrt{\delta} - \sqrt{\delta-8} \pm \sqrt{10\delta+22\sqrt{\delta(\delta-8)}-8}
\Bigr)$.
\begin{figure}
\includegraphics[angle=0,width=1.52in]{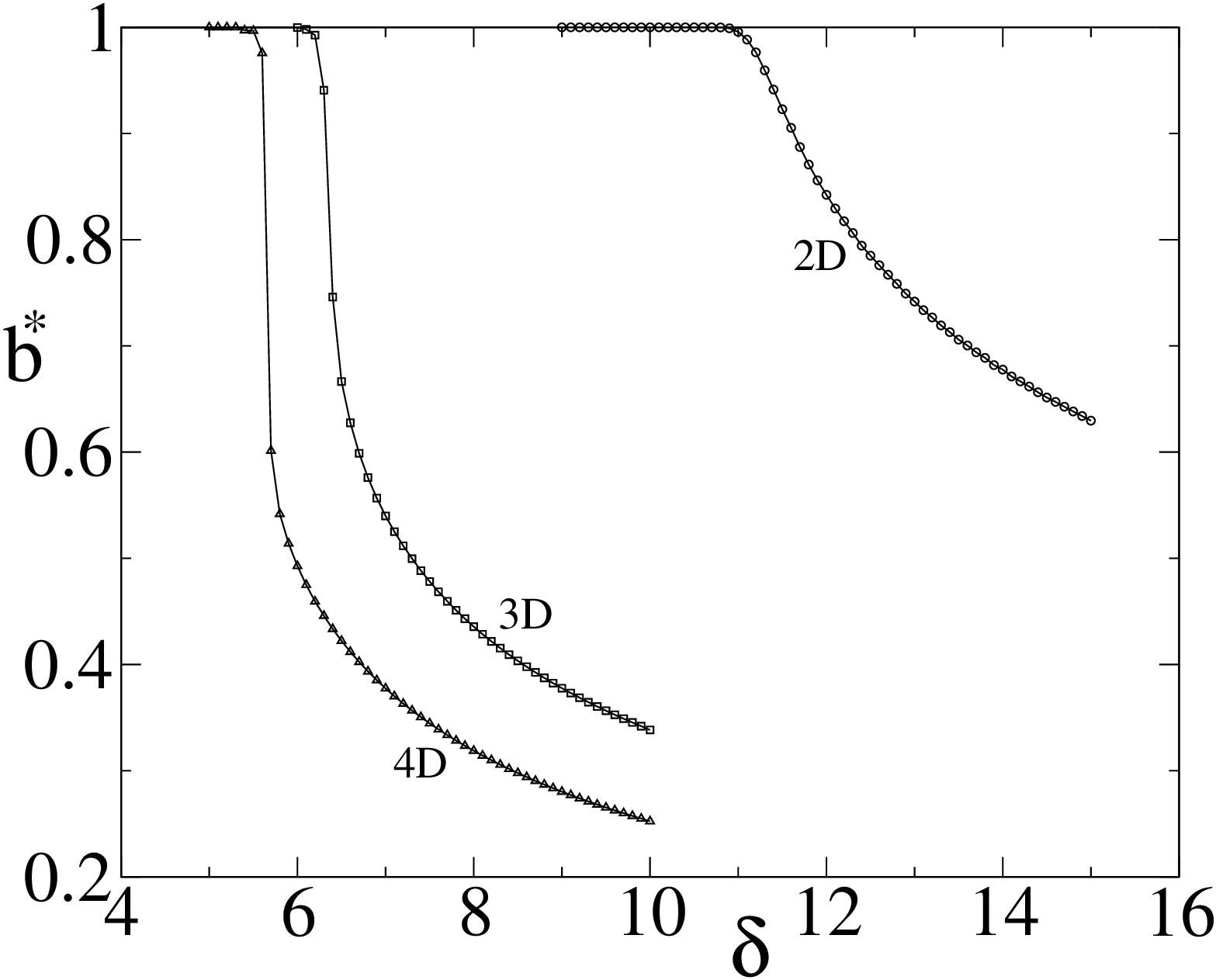}
\includegraphics[angle=0,width=1.64in]{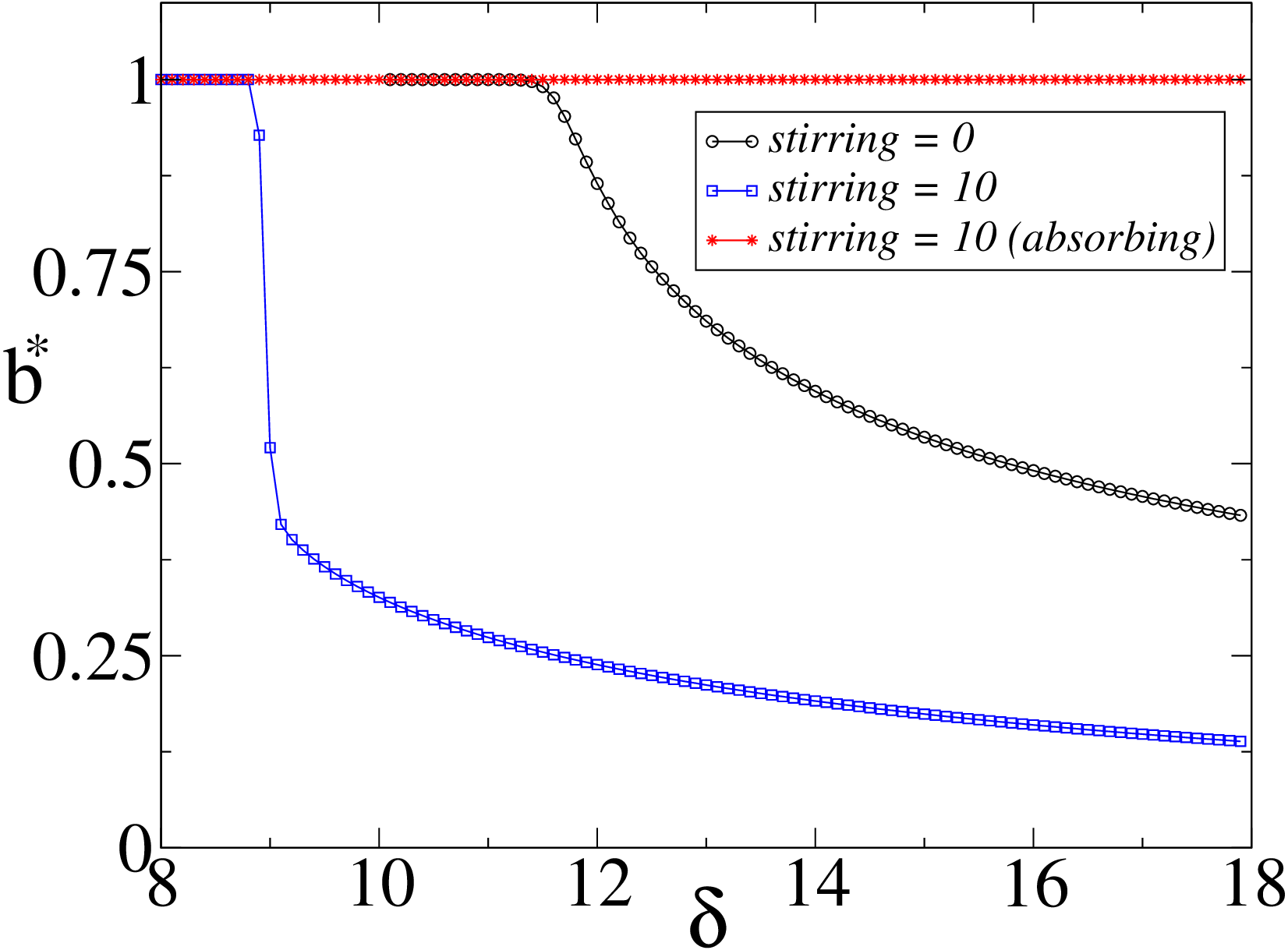}
\caption{(Color online)\, Average stationary prey density $b^*$ vs $\delta$ for
$\eta = \sigma = 2\mu = 2$.
Left: DP-like transitions on  $256^2, 50^3$ and $20^4$ lattices when
${\cal D} = 0$. Right: Effect of the stirring on a $512^2$ lattice.
For ${\cal D} = 0$ there is a {\em continuous} transition (center curve, black),
while for ${\cal D} = 10$ (sufficient stirring) two stable branches emerge and 
there is a {\em first-order} transition; the top (red) branch corresponds to 
predator extinction, and the bottom (blue) one is associated with a 
coexistence phase.}
\label{NNN_trans}
\end{figure}
It follows from this discussion that, at the mean-field level, the introduction
of a triplet interaction changes the behavior of the system dramatically:
For $\delta > \delta_c$ the system can reach the absorbing state full of prey
or alternatively a phase where the prey population, with stationary density
$b^* < 1/2$, coexists with the predators.
Hence, the rate equations predict the possibility of a
{\em first-order} phase transition.

\begin{figure}
\includegraphics[angle=0,width=1.80in]{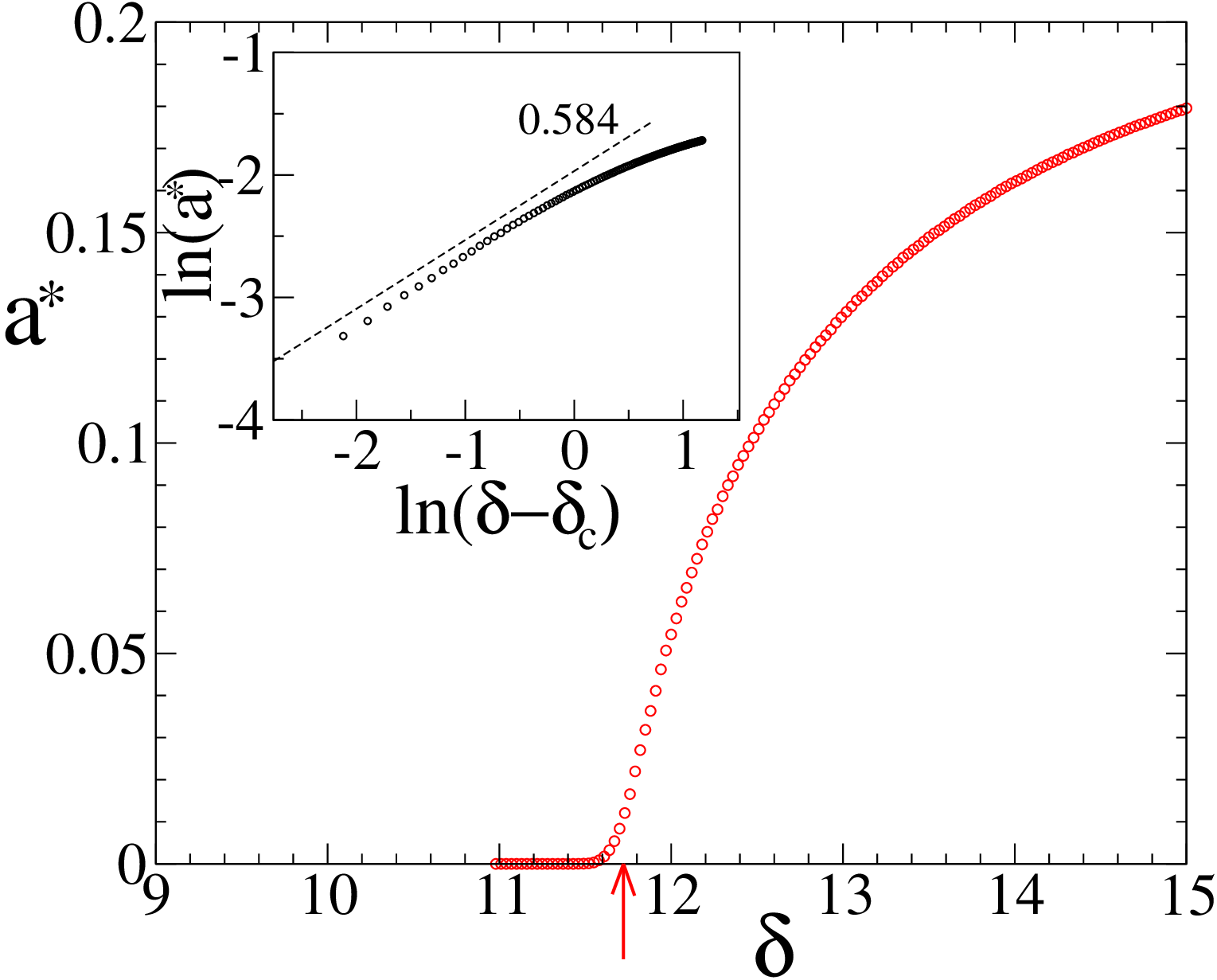}
\caption{(Color online)\, Average stationary density of predators in the 
absence of stirring on a $512 \times 512$ lattice with 
$\eta = \sigma = 2\mu = 2$ and ${\cal D} = 0$: existence of a DP-like phase 
transition at $\delta_c \approx 11.72$ with exponent $\beta \approx 0.584$ 
(see inset).}
\label{NNN_DP}
\end{figure}

Motivated by these predictions, quite different from those of other LPPM, we
have studied the properties of our NNN-LPPM through Monte Carlo simulations on
periodic hypercubic lattices.
We have first considered the case of slow (${\cal D} \approx 0$) and fast
stirring and noticed the emergence of quite different behavior.
In fact, for no (or slow) stirring, instead of a discontinuous phase
transition, we have observed a {\em continuous} active-to-absorbing phase
transition as for the LPPM in dimensions $d = 2$, $3$ and even $d=4$, see
Fig.~\ref{NNN_trans} (left). 
(Of course, in dimensions $d > 3$ the model is biologically irrelevant: 
these cases have only been considered to assess the validity of the MF theory.)
To ascertain the properties of the NNN-LPPM we have employed the dynamical
Monte Carlo technique \cite{Hinrichsen}.
Near the extinction threshold, one expects power law behavior for the survival
probability $P(t) \sim t^{-\delta'}$ and the number of active sites
$N(t) \sim t^{\theta}$.
By averaging over $3 \times 10^6$ independent runs, performed on a
$512 \times 512$ lattice, each with duration $10^5$ Monte Carlo steps, for
fixed rates $\eta = \sigma = 2\mu = 2$, and ${\cal D} = 0$ we have estimated 
the critical point to be at $\delta_c \approx 11.72$ (the MF prediction is
$\delta_c = 8$), and measured $\delta' \approx 0.451$ and
$\theta \approx 0.230$, very close to the established two-dimensional DP
exponents \cite{Hinrichsen}.
As illustrated in Fig.~\ref{NNN_DP}, we have also determined the order
parameter critical exponent defined via
$a(t \to \infty) \sim (\delta - \delta_c)^\beta$ as $\beta \approx 0.584$.
We have checked that the exponent values are consistent with the DP
universality class for several choices of the rates $\eta,\mu,\sigma$.
Qualitatively, the features of the NNN-LPPM remain similar in $d=3$ and $4$ 
(Fig.~\ref{NNN_trans}, left): we observe continuous phase transitions (for 
different values of $\delta_c$) with $\beta \approx 0.81$ for $d = 3$ and 
$\beta \approx 1.0$ for $d = 4$ (upper critical dimension of DP) in 
agreement with DP values \cite{Hinrichsen}.

In the absence of stirring, the phase diagram changes qualitatively when
$d \geq 5$: even for ${\cal D} = 0$, one then observes the first-order phase
transition predicted by the MF approximation.
The situation turns out to be completely different when the stirring is
sufficiently fast, as illustrated in Fig.~\ref{NNN_trans} (right):
A first-order phase transition occurs in low dimensions as well,
and, depending on the initial condition with respect to the separatrices
(see Fig.~\ref{portrait}, right), the flows in the phase portrait end either at
the absorbing fixed point $(0,1)$, or reach a stationary
state where both predators and prey  coexist (with  $b^* < 1/2$).
This scenario, in the presence of sufficiently fast stirring, therefore recovers
the {\em mean-field} behavior, at least qualitatively.
It is quite remarkable that the rate equations (\ref{NNN_MFa}), (\ref{NNN_MFb})
describe the NNN-LPPM already for mere {\em nearest}-neighbor (NN) exchanges at
{\em finite} rates; one would rather expect the MF regime to emerge in
the limit of infinitely fast exchange processes involving the swap of {\em all}
particles (not restricted to NN partners) \cite{Durrett}.

\begin{figure*}
\includegraphics[width=1.6in]{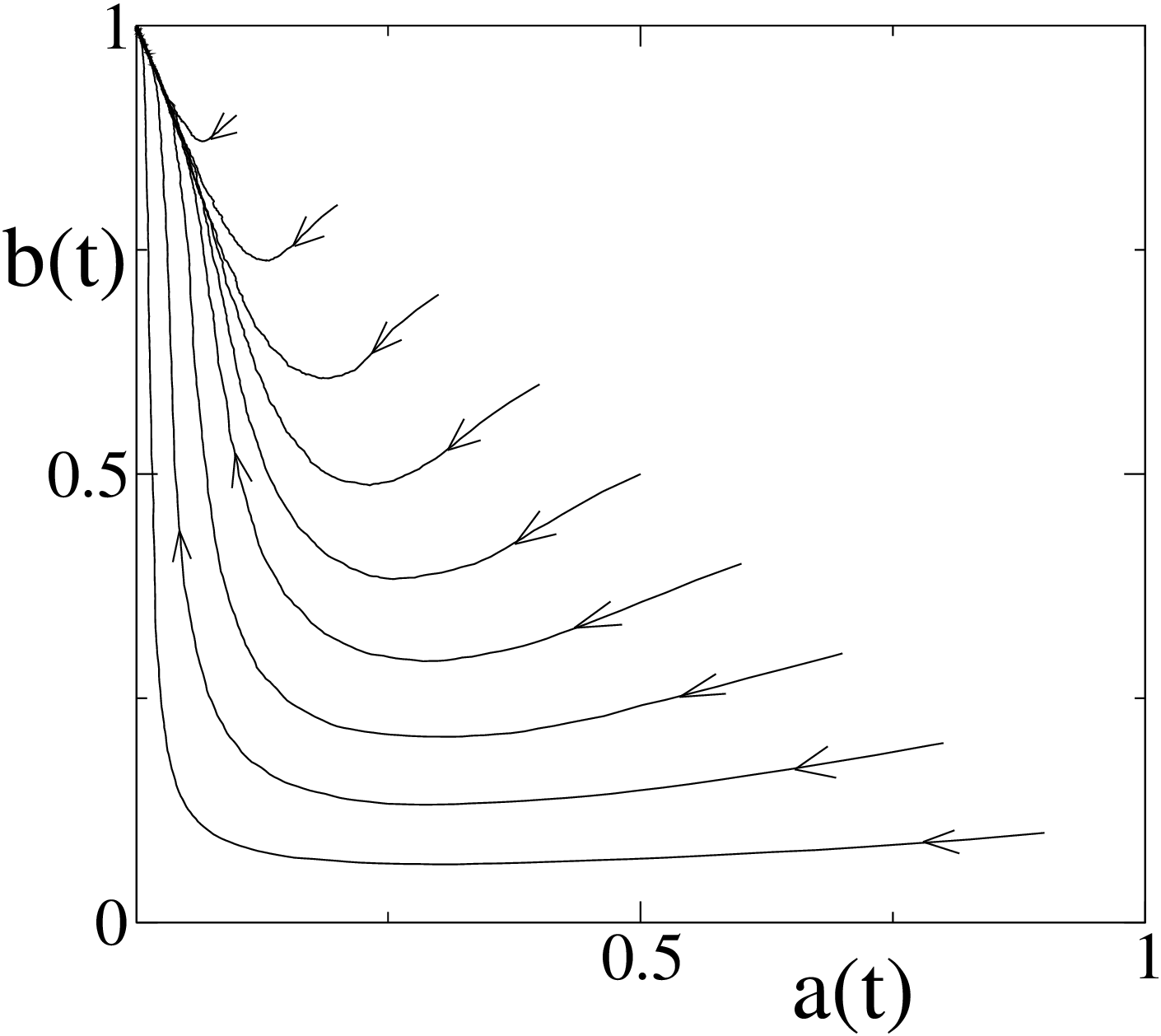}
\includegraphics[width=1.62in]{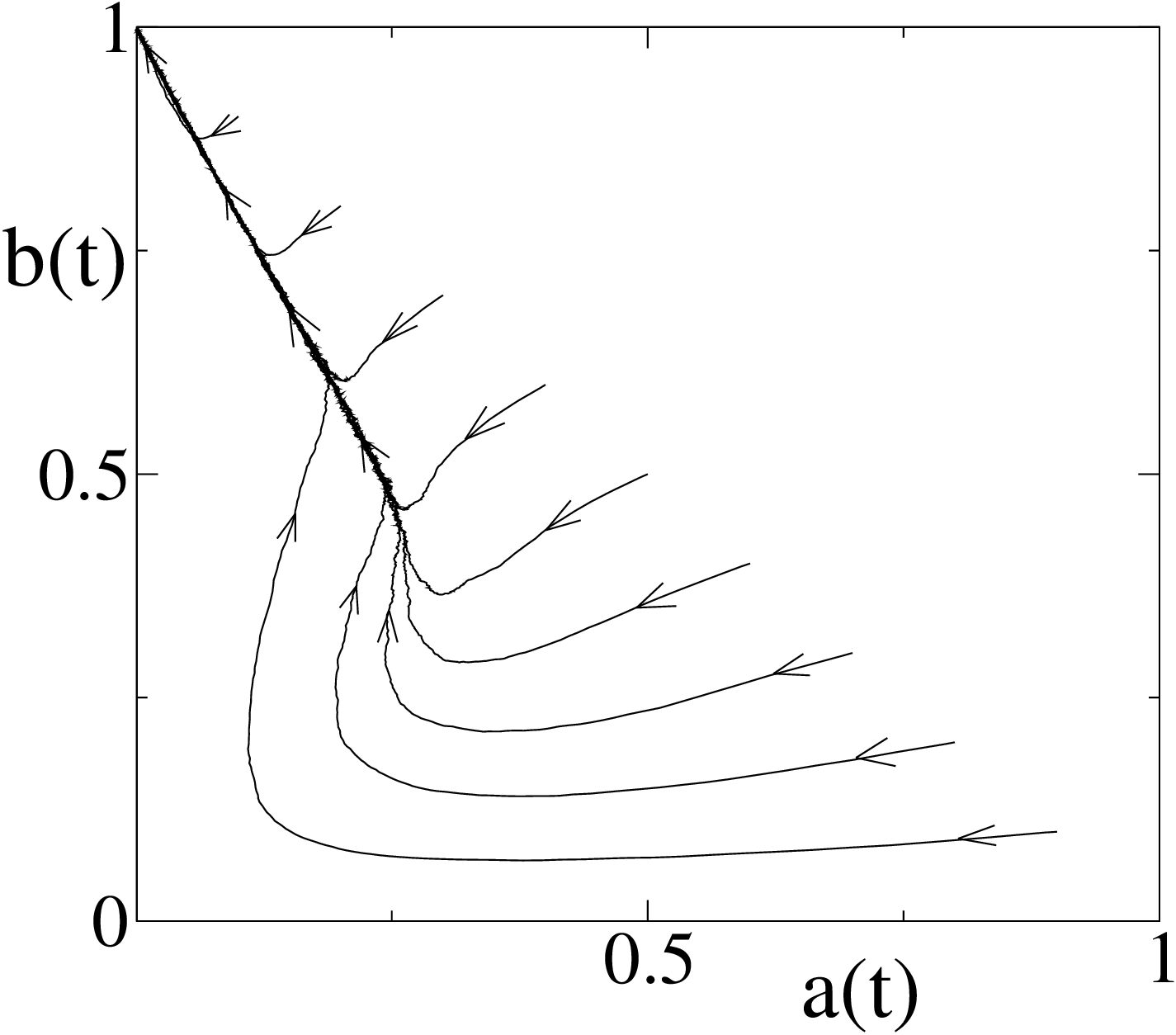}
\includegraphics[width=1.6in]{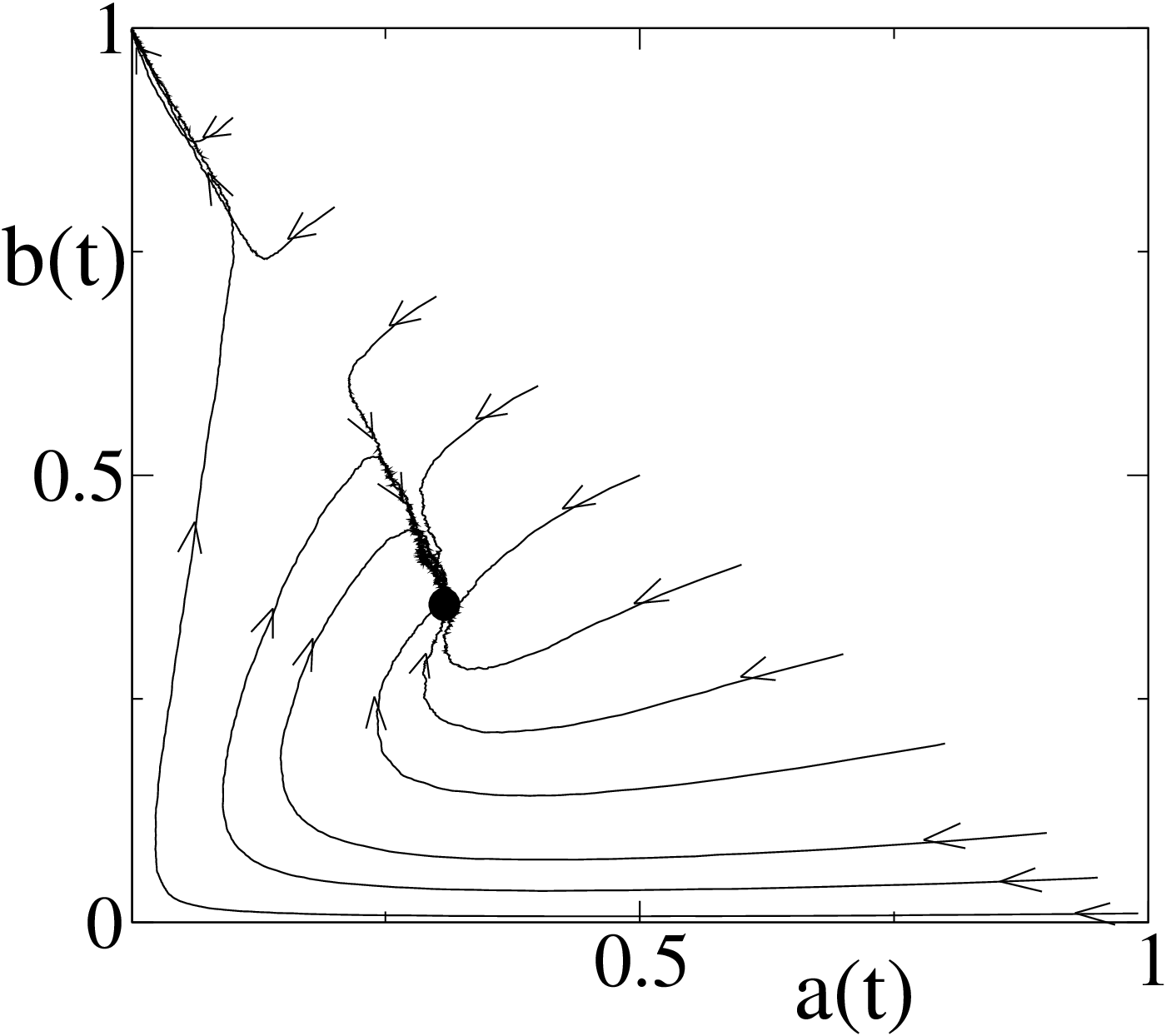}
\caption{Phase portrait of the generalized `NNN' lattice predator-prey
model (on a $256 \times 256$ lattice) with rates $\eta = \mu = \sigma = 1$,
$\delta = 10$ and different exchange rates.
>From left to right: `stirring' rate ${\cal D} = 0, 2, 5$. (See text).}
\label{2d_stir}
\end{figure*}

As illustrated in Fig.~\ref{2d_stir}, the intriguing properties of the 
NNN-LPPM with NN exchange process can be summarized as follow:
(i) For vanishing mixing (${\cal D}$ small compared to the other rates), in
dimensions $1 < d \leq 4$ the system undergoes an active-to-absorbing state
transition which belongs again to the DP universality class; only for
$d \geq 5$, a first-order phase transition appears.
Stochastic fluctuations clearly have a drastic effect here, invalidating the
mean-field picture in dimensions $d \leq 4$.
(ii) When one allows for random short-range particle mixing (${\cal D} > 0$),
the dynamics and the phase portrait flows change dramatically 
[Fig.~\ref{2d_stir}, center].
(iii) When the exchange processes become sufficiently fast (typically, when
${\cal D} \approx \delta$) a new fixed point associated with a coexistence
phase is available (this holds even in $d = 1$), as demonstrated in
Fig.~\ref{2d_stir} (right), and the system undergoes a first-order phase transition
as predicted by the mean-field theory.
As expected, when there is `fast' stirring (${\cal D}$ much larger than the 
other rates) the MF predictions become very accurate.
We have also checked that the NNN-LPPM stable active fixed point is, in 
agreement with the MF analysis and generic properties of the other LPPM 
\cite{Lipowska,Provata}, either a node or a focus. 
When it is a focus, the coexistence phase is again characterized by population 
oscillations originating in moving activity fronts but, as the system is 
more mixed, these `rings' appear less prominent than in the LPPM with NN interactions \cite{Mobilia}.

In this paper, we have first outlined 
the main properties of the LPPM
with nearest-neighbor interactions: namely, the existence of 
erratic oscillations and complex patterns deep in the coexistence phase and a 
directed percolation type phase transition.
We have then further tested this scenario by considering a perhaps more
realistic model variant with \emph{next-nearest-neighbor} interaction. 
Upon in addition introducing a short-range stirring mechanism together with
this longer-range interaction, an intriguing interplay emerges: When the 
NN exchange process is `slow', the NNN reaction induces subtle correlations that
completely invalidate the MF treatment and the system still undergoes a DP-type
phase transition (for $1 < d \leq 4$).
In this regime, the generic LPPM scenario is thus fully confirmed.
However, when the value of the mixing rate ${\cal D}$ is raised, the simple NN 
exchange process `washes out' the NNN correlations and the system reproduces 
the MF behavior, displaying a \emph{first-order} phase transition.
This is to be viewed in contrast with the standard LPPM, for which even fast 
diffusion of predators and prey generally does not qualitatively affect its 
properties \cite{Lipowska}.

\noindent
We gratefully acknowledge the support by the U.S. NSF under grants
DMR-0088451, 0308548, 0414122, and (M.M.) by the Swiss NSF and 
the Humboldt Foundation (grants 81EL-68473 and IV-SCZ/1119205 STP).
We thank T.~Antal, J.~Banavar, R.~Bundschuh, E.~Frey, P.L.~Krapivsky,
R.~Kulkarni, T.~Newman, G.~Pruessner, B.~Schmittmann, N.~Wingreen, and
R.~K.~P.~Zia for inspiring discussions.

\end{document}